# Joint Geometry and Color Projection-based Point Cloud Quality Metric

A. Javaheri, C. Brites, *Member,* IEEE, F. Pereira, *Fellow,* IEEE, J. Ascenso, *Senior Member,* IEEE

*Abstract* — **Point cloud coding solutions have been recently standardized to address the needs of multiple application scenarios. The design and assessment of point cloud coding methods require reliable objective quality metrics to evaluate the level of degradation introduced by compression or any other type of processing. Several point cloud objective quality metrics has been recently proposed to reliable estimate human perceived quality, including the so-called projection-based metrics. In this context, this paper proposes a joint geometry and color projection-based point cloud objective quality metric which solves the critical weakness of this type of quality metrics, i.e., the misalignment between the reference and degraded projected images. The best performing 2D quality metrics in the literature are exploited to evaluate the quality of the projected images. The experimental results show that the proposed projection-based quality metric offers the best subjective-objective correlation performance in comparison with other metrics in the literature. The Pearson correlation gains regarding D1-PSNR and D2-PSNR metrics are 17% and 14.2% when data with all coding degradations is considered.**

## I. Introduction

RECENT advances in 3D acquisition and reconstruction technologies have enabled many visual immersive applications, such as virtual and augmented reality, immersive communications, and video gaming. Point cloud (PC) is an emerging 3D visual representation format that is becoming rather popular because of its easy acquisition and capability to realistically represent objects and 3D visual scenes. However, since realistic PCs require a large number of points, a compact representation of PCs is essential for storage and transmission applications and services. PC coding is a rather new and challenging problem due to the unstructured nature of PCs where each point, i.e., filled voxel in voxelized PCs, is associated to a 3D coordinate; moreover, each point has often associated attributes such as color, transparency, reflectance, etc. In recent years, several efforts in PC coding were able to significantly reduce the bitrate while still maintaining the PC quality and fidelity. MPEG has already developed two PC coding standards [1]-[3], notably Geometry-based Point Cloud Compression (G-PCC) for static and progressive acquired PCs, and Video-based Point Cloud Compression (V-PCC) for dynamic PCs. In this context, the assessment of PC quality is very important as it plays a significant role in the design and optimization of coding solutions as well as on the validation of the quality of experience offered to the users.

The best way to reliably measure the PC quality is through subjective quality assessment where a specially designed framework collects opinion scores from a minimum number of subjects. In the literature, there are many subjective PC quality studies available, considering different ways to visualize the PC [4][5], several PC rendering methods [6][7] and types of degradation [8][9]. However, since subjective quality assessment is expensive and time-consuming, reliable objective quality metrics are critical to facilitate the design of more efficient coding solutions and assess the quality of experience offered to the users. In the literature, the performance of multiple PC objective quality metrics has been assessed through the correlation between corresponding objective and subjective quality scores, notably for different codecs, with different types of degradation [6][10].

In the literature, a few works have already exploited the idea of measuring the PC quality by projecting the 3D PC into one or more 2D images, i.e., by converting a 3D PC into several 2D images, a more traditional type of data. In the context of PC coding, this type of approach was successfully exploited to achieve efficient compression, as demonstrated by the MPEG V-PCC standard [2][3]. In the context of PC quality assessment, these 2D projected images can be obtained by performing multiple projections into different viewpoints, i.e., using different projection centers. After projection, the most recent, and powerful 2D image quality metrics available can be exploited without any changes, to assess the entire PC quality through the projected images quality. However, the projection-based metrics available in the literature are not yet showing better subjective-objective correlation performance than the popular point-to-point quality metrics, where correspondences are established in the 3D space and errors/distances in position or color are accounted.

The most critical weakness of projection-based metrics is caused by the inability of 2D quality metrics to efficiently handle local displacement errors, since pixel-level (or region-level) comparisons are usually made. Due to lossy PC coding, geometry distortions (or degradations) cause many displacements and thus a lower correlation performance. Typically, 2D objective quality metrics consider that these pixels/regions have a high distortion when, in fact, the small or medium geometry degradations are perceptually well tolerated, especially when color is also available as some degree of masking may happen [6]. For example, small displacement errors in the projected images due to geometry distortions may not be perceived by humans but may lead to high objective distortions when 2D quality metrics are used to assess the quality of the PC projected images. Another critical issue is related to the difference between the number of points in the reference and decoded PCs. This often occurs when the PC coding solution uses planar or triangular approximations of the PC surface, and more points may be recreated at the decoder side when these surfaces are sampled or when the PC coding solution reduces the number of coded points using octree pruning. This implies that one of the projected 2D images, the reference or the degraded one, may have, for some positions, pixels occupied while these pixels are not filled in the other projected 2D image, thus leading to large pixel-based mismatches. In some past works [11] [12], these positions are either ignored or an occupied position is compared with a non-occupied position (usually filled with some background color). However, both these solutions negatively impact the final quality metric correlation performance since, for some cases, these pixels are visually important and should not be ignored; moreover, the quality score should not depend on an arbitrarily selected background color.

In this context, this paper proposes a novel joint geometry and color projection-based PC quality metric, which addresses the weaknesses and issues above, thus allowing to achieve a higher objective-subjective correlation performance. The key original ideas underpinning this novel quality metric are twofold:

- Reference and degraded projected images are compared for two fixed geometry conditions, notably for the reference and degraded geometries. After, these two quality scores are fused, thus implying that the proposed approach implicitly considers geometry and color distortions/degradations. By comparing images created for the same geometry level, reference or degraded, the undesired above misalignments are avoided and there is no difference between the number of points on the reference and degraded PCs for the same geometry condition.

- A padding operation is performed in the 2D domain to avoid assigning an arbitrary, uniform color to the background (i.e., not projected) pixels. Because of the aligned geometries, these pixels are not filled in the reference/degraded projected images and if a uniform background value is assigned, the 2D quality metric would be biased due to these regions for which no distortions would occur. The proposed padding operation mitigates the impact of these background pixels.

To achieve its objectives, the rest of the paper is organized as follows. Section II briefly reviews the state-of-the-art on PC objective quality metrics. Section III describes the proposed joint geometry and color projection-based PC quality metric. Experimental results are presented and analyzed in Section IV and, finally, conclusions are offered in Section V.

## II. BACKGROUND WORK ON PC OBJECTIVE QUALITY ASSESSMENT

In this section, the state-of-the-art on PC objective quality metrics is briefly reviewed, by addressing first, point-based metrics, followed by feature-based metrics and, finally, projection-based metrics.

### A. Point-based PC Quality Metrics

A point-based quality metric compares the geometry or attributes of the reference and degraded PCs directly point-by-point after defining the necessary point correspondences. The most popular point-based geometry quality metrics are the Point-to-Point (Po2Po) [13] and Point-to-Plane (Po2Pl) [14] metrics. In a Po2Po metric, for every point in a degraded/reference PC, the nearest neighbor is obtained in the corresponding reference/degraded PC (thus a point correspondence is obtained); after, the Hausdorff distance or the Mean Squared Error (MSE) distance are computed over all pairs of points. The main disadvantage of this type of metrics is that they do not consider that PC points represent the surface of an object(s) in the visual scene. To solve this issue, Point-to-Plane (Po2Pl) metrics have been proposed by Tian *et al.* [14], which model the underlying surface at each point as a plane perpendicular to the normal vector at that point. This type of metrics results into smaller errors for the points closer to the PC surface, here associated to a plane. Currently, the MPEG-adopted PC geometry quality metrics are the Po2Po MSE (D1) and Po2Pl MSE (D2) distances/errors and their associated PSNR [15]. Moreover, a Plane-to-Plane (Pl2Pl) metric has been proposed by Alexiou *et al.* [16], which measures the similarity between the underlying surfaces associated to the corresponding points in the reference and degraded PCs. In this case, tangent planes are estimated for both the reference and degraded points and the associated angular similarity is assessed.

In [17], Javaheri *et al.* propose a geometry quality metric based on the Generalized Hausdorff distance, which corresponds to the maximum distance for a specific percentage of data rather than the whole data, thus filtering some outlier points. The Generalized Hausdorff distance between two PCs adopted in this quality metric may be computed for both Po2Po and Po2Pl metrics. In [18], Javaheri *et al.* also propose a so-called Point-to-Distribution (Po2D) metric based on the Mahalanobis distance between a point in a PC and its *K* nearest neighbors in the other PC. The mean and covariance matrix of the corresponding distribution are computed and used to measure the Mahalanobis distance between points in one PC and their corresponding set of nearest neighbors in the other PC. These distances are averaged to compute the final quality score.

There are not many point-based quality metrics for PC attributes and specifically for color. However, the Po2Po PSNR for color in the $YC_bC_r$ color space is widely used by MPEG and in the literature to evaluate PC color degradations. This metric works like the Po2Po geometry metrics, with the error corresponding now to the difference between the colors of the points in some PC correspondence. This metric may either be computed only for the luminance (Y-PSNR) or chroma components ($C_b$/$C_r$-PSNR) individually, or as a weighted average of all color components (YUV-PSNR).

### B. Feature-based PC Quality Metrics

A feature-based PC quality metric computes a quality score based on the difference between some local or/and global features extracted from the reference and degraded PCs. Meynet *et al.* propose in [20] the so-called Point Cloud Multi-Scale Distortion metric (PC-MSDM), a structural similarity-based PC geometry quality metric based on local curvature statistics. This metric computes the surface curvature associated to each point and establishes after point-based correspondences. The metric score corresponds to the Gaussian weighted curvature statistics for a set of local neighborhoods.

In [21], Viola *et al.* design a PC quality metric based on the histogram and correlogram of the luminance component. After, the proposed color quality metric is fused with the Po2Pl MSE geometry metric (D2) using a linear model with a weighting parameter determined using a grid search method.

In [22], Diniz *et al.* propose the so-called Geotex metric, which is based on Local Binary Pattern (LBP) descriptors adapted to PCs and applied to the luminance component. To apply it on PCs, the LBP descriptor is computed on a local neighborhood corresponding to the $K$-nearest neighbors of each point in the other PC. Histograms of the extracted feature maps are obtained for both the reference and degraded PCs and used to compute the final quality score using a distance metric such as the f-divergence [23]. In [24], Diniz *et al.* extend the Geotex metric by considering multiple distances, notably Po2Pl MSE for geometry and the distance between LBP statistics [22] for color. In [25], Diniz *et al.* also propose another quality metric, which computes Local Luminance Patterns (LLP) on the $K$-nearest neighbors of each point on the other PC.

In [26], Meynet *et al.* propose the Point Cloud Quality Metric (PCQM) metric, which combines the geometry features used in [20] with five color features related to lightness, chroma and hue. PCQM corresponds to the weighted average of the differences for geometry and color features between the reference and degraded PCs. In [27], Viola *et al.* propose the first reduced reference quality metric that jointly evaluates geometry and color. A set of seven statistical features such as the mean and standard deviation are extracted from the reference and degraded PCs in the geometry, texture, and normal vector domain, in a total of 21 features. The reduced quality score is computed as the weighted average of the differences for all these features between the reference and degraded PCs.

Inspired by the SSIM quality metric for 2D images, Alexiou *et al.* propose in [28] a quality metric using local statistical dispersion features. These statistical features are extracted in a local neighborhood around each point in the reference and degraded PCs considering four independent 'attributes', notably geometry, color (luminance), normal and curvature information. The final quality metric is obtained by pooling the feature value differences between associated points in the reference and degraded PCs.

### C. Projection-based PC Quality Metrics

A projection-based PC quality metric maps the 3D reference and degraded PCs onto some selected 2D planes and computes the quality score by comparing the projected images using some 2D image quality metric. The first projection-based PC quality metric has been proposed by Queiroz *et al.* in [11]. This metric projects the reference and degraded PCs onto the six faces of a bounding cube around the PC, concatenates the corresponding projected images and measures the 2D PSNR between the corresponding degraded and reference concatenated projected images.

In [12], Alexiou *et al.* propose a rendering software for PC visualization on 2D screens, which performs the orthographic projection of a PC onto the six faces of the PC bounding box. A 2D quality metric is then applied to the reference and degraded projected images resulting from the rendering and the final score is obtained by computing the average over the six pairs of projected images. In [29], Alexiou *et al.* study the impact of the number of projected 2D images (each corresponding to a specific view) on the subjective-objective correlation performance of projection-based quality metrics. It is shown that even a single view may be enough to achieve a reasonable correlation performance. Moreover, a projection-based PC quality metric weighting the projected images according to the user interactions performed during the subjective test is proposed. In [10], the quality metric proposed in [12] is benchmarked considering different number of views, pooling functions, etc. The best performance is achieved when 2D quality metrics are applied on the projections from 42 different views and pooled with an l1-norm.

## III. PROPOSED PROJECTION-BASED PC QUALITY METRIC

In this section, the architecture and walkthrough of the proposed joint geometry and color projection-based PC quality metric is presented; after the most relevant modules are explained in detail.

### A. Architecture and Walkthrough

Fig. 1 shows the proposed Joint Geometry and Color Projection-based PC Quality Metric architecture, referred from now on as *JGC-ProjQM*. The key idea behind this metric is that the degraded and reference PCs are processed in two parallel branches, one associated to the reference geometry and the other associated to the degraded geometry, to obtain two intermediate quality scores which are fused at the end. To avoid misalignment errors, before applying the 3D to 2D projection, the reference and degraded PCs are processed to obtain two PCs to be compared with the same geometry:

- **Reference Geometry Branch:** In the top branch, the geometry of the reference PC is used; the reference PC geometry is recolored with the color of the degraded PC and the resulting PC is compared with the reference PC (naturally, including the original color).
- **Degraded Geometry Branch**: In the bottom branch, the geometry of the degraded PC is used; the degraded PC

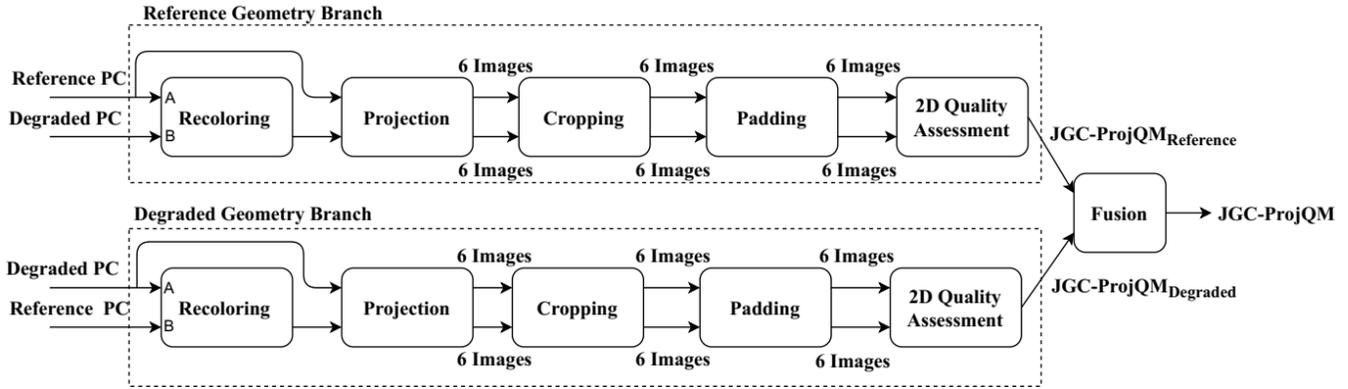

Fig. 1. Proposed joint geometry and color projection-based PC quality metric (*JGC-ProjQM*) architecture.

geometry is recolored with the reference PC color and the resulting PC is compared with the degraded PC (naturally, including the degraded color).

Naturally, the color attributes assigned to the points in the two PCs with the same geometry in the two branches will be different, notably using the color data with and without coding degradations.

Before applying the *JGC-ProjQM* metric, the PCs have to be voxelized to some fixed precision, if they are available in floating point precision. This step is important to perform the 3D (voxels) to 2D (pixels) projection. Nowadays, most available PC data are already in fixed precision, i.e., the PC has been voxelized, and, thus, this step may not be needed; for this reason, it is not included in the architecture in Fig. 1. Moreover, both V-PCC and G-PCC standard codecs code fixed precision PCs or perform voxelization as a pre-processing step before coding. The main modules in the proposed *JGC-ProjQM* metric pipeline are briefly explained in the following walkthrough:

1. **Recoloring:** Due to the lossy coding of geometry data, the positions of the points in the reference and degraded PCs are not the same. Thus, when a projection is made, the resulting reference and degraded projected 2D images have regions that are not aligned, even when the degraded PC has only slight geometry degradations. This creates a problem for 2D quality metrics which are typically not robust to misalignments (or displacements) and often perform very poorly in this situation. Therefore, to align the reference and degraded projected images, a recoloring step is applied where the degraded (or reference) PC color is mapped onto the reference (or degraded) geometry. In this way, the color degradation is compared for the two geometry conditions (reference and degraded geometry, each at a time), without any misalignments (or different number of points) while still considering both the geometry and color degradations. This is a key technical novelty of the proposed *JGC-ProjQM* quality metric. This solution also exploits the fact that color degradations typically have a higher perceptual impact than geometry degradations, notably due masking effects [6]. More details about this module are presented in Section III.B.

2. **Projection:** The reference, degraded and the two recolored PCs obtained in the previous step are orthographically projected onto the six faces of a cube to obtain six projected images for each PC, this means six non-overlapping images for each of the four PCs, i.e., reference, degraded, reference recolored and degraded recolored. Although another type of projection could be used, the low-complexity orthographic projection is enough to assess the quality degradations, especially considering that the PC pairs to compare have now the same geometry. In this process, six binary occupancy maps are also obtained for each projected PC; these occupancy maps serve to signal if a 2D image pixel corresponds (or not) to a point (filled voxel) in the 3D PC. The size of the projected images and occupancy maps only depends on the precision $p$ of the PC, thus commonly obtaining a $2^p \times 2^p$ size. More details about this module are presented in Section III.C.

3. **Cropping:** After the projection, and depending on the PC size and position, the projected images may have a rather large background area (area without projected pixels around the PC object(s)), notably in comparison with the image area occupied with the PC points. These empty background areas can act as a distractor for the 2D quality metric, notably if the same (uniform) color value is assigned to all background pixels and thus they must be reduced as much as possible with a cropping procedure. More details about this module are presented in Section III.D.

4. **Padding:** If an arbitrary, uniform background value is used for background pixels, i.e., pixels positions on the projected images that have not been filled, the 2D quality metric may be biased, thus lowering the *JGC-ProjQM* prediction power. Computing the 2D quality metric only on the foreground pixels is not also a good solution as some 2D quality metrics such as SSIM cannot be applied to arbitrarily shaped objects. In this context, the padding module targets the creation of a seamless image, where the background positions are filled with interpolated/padded values, thus obtaining an image that is more suitable for quality assessment. In this process, the background holes inside the PC foreground area are also padded in the same way as the empty areas around the projected PC. More details about this module are presented in Section III.E.

5. **2D Quality Assessment:** At this stage, a 2D image quality metric is computed between the six reference, padded images and the corresponding degraded, padded images for the same view/projection; this happens for the two architectural branches. The output of this process are six objective quality scores, one for each pair of projected, padded images, corresponding to each projection plane, which must be pooled together. The final *JGC-ProjQM* metric performance has been studied for several 2D quality metrics, using common pooling functions, e.g., max, min, and weighted average. Since it was found that the final subjective-objective correlation performance is rather similar for the various pooling functions, it was decided to adopt average pooling to obtain a single quality score for each architectural branch of the proposed projection-based PC quality metric.

6. **Fusion:** All modules previously described are included in the two architectural branches of the proposed projection-based PC quality metric to obtain two intermediate quality scores, notably *JGC-ProjQM_reference* and *JGC-ProjQM_degraded*. These two intermediate quality scores represent the quality associated to the projected images as measured by a 2D quality metric, for two different geometry conditions, i.e., reference and degraded geometries, and must be fused together to obtain the final *JGC-ProjQM* quality metric. Although different fusion strategies are possible, even applying machine learning techniques, it was found that a simple linear regression was enough to obtain a high subjective-objective correlation performance, without the risk of overfitting. More details about this process are presented in Section III.F.

### B. Recoloring

The main challenge with projection-based PC quality metrics is that geometry distortions may cause misalignment errors between the reference and degraded projected 2D images. 2D quality metrics do not typically handle well local displacement errors; for example, when the same pixel location in images projected for reference and degraded geometry is compared, the measured error may not express well the user perceived distortion since the color of different 3D positions in the reference and degraded PCs are used. Fig. 2 shows the frontal projection for the *Egyptian Mask* PC before coding (reference geometry and color) and after G-PCC decoding and recoloring with the original/reference color, for the lowest geometry rate. Fig. 2 also shows a residual image with the difference between the previous reference and decoded, recolored projected images (with some enhancement for better visualization). Although the color in both PCs is the same and the PCs are visually similar, the residual image shows large misalignment errors.

To overcome this key problem, this paper proposes an original solution involving computing the 2D quality metrics with different color data under two geometry conditions, notably reference and degraded geometries. The idea is to use the PC geometry, reference and degraded, and to perform recoloring to assign the decoded color to the reference geometry (top branch of the architecture) and the reference color to the decoded geometry (bottom branch of the architecture). By using this approach, the projected images are always geometry-aligned within each branch, thus avoiding misalignments.

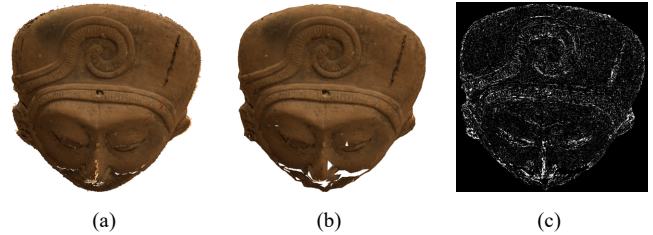

Fig. 2. *Egyptian Mask* projected from front view: (a) Reference PC; (b) Decoded G-PCC geometry for lowest geometry rate, recolored with the original color; (c) Residual image between (a) and (b) after enhancement.

To recolor PC *A*, with the color of PC *B*, each point in PC *A* will have a color assigned using the color of one or more corresponding points in PC *B*. In the proposed recoloring algorithm, the color for each point in PC *A* after recoloring is determined as follows:

1. For each point in PC *A*, the nearest neighbor in PC *B* is found ($NN_A$) and, for each point in PC *B*, the nearest neighbor in PC *A* is found ($NN_B$).

2. For each point *a* in PC *A* perform: if point *a* is listed in the nearest neighbors of some points in PC *B* ($a \in NN_B$), its color is the average color of the points in PC *B* which have point *a* as their nearest neighbor, as defined in (1):

$$C_a = \sqrt{\frac{\sum_{b \in B, NN_B(b)=a} C_b^2}{\sum_{b \in B, NN_B(b)=a} 1}} \qquad (1)$$

Otherwise, its color is the color of its nearest neighbor listed in $NN_A$. In (1) $C_a$ and $C_b$ are the colors at points *a* in PC *A* and *b* in PC *B*. The denominator counts the number of points in PC *B* which have point *a* in PC *A* as its nearest neighbor.

Fig. 3 illustrates the recoloring process for the *Amphoriskos* PC using a point-based rendering solution with cube primitives. Fig. 3(a) shows the reference PC (reference geometry and color) on the left and the recolored PC (reference geometry and decoded color) on the right. Fig. 3(b) shows the degraded PC (decoded geometry and color) on the left and the recolored PC (decoded geometry and reference color) on the right. Each pair of PCs have the same geometry and either the reference or degraded colors; because of the geometry alignment, it is now possible to compare these color values using a simple, direct pixel-to-pixel correspondence.

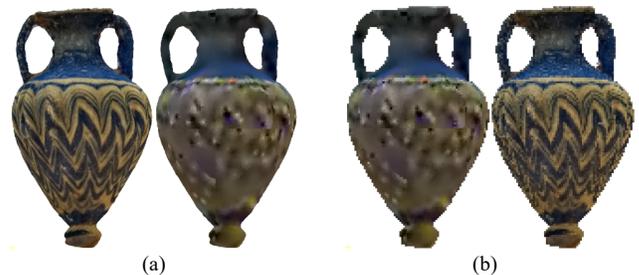

Fig. 3. *Amphoriskos* PC decoded with G-PCC in Octree geometry coding mode and Lifting color coding mode at the lowest rate: (a) PCs obtained for the reference geometry with the reference (left) and decoded color, after recoloring (right); (b) PCs obtained for the decoded geometry with the decoded color (left) and reference color, after recoloring (right).

### C. Projection

This is a core module of the proposed *JGC-ProjQM* architecture, where 3D PCs are mapped onto six 2D planes from different perspectives, thus creating the projected images. The proposed projection procedure is based on the orthographic projection, an often-used parallel projection that renders objects with suitable shapes and sizes. In this procedure, each PC point is projected to a pixel in a 2D image while considering its visibility (or occlusion) for each specific viewing perspective and, thus, projection plane. The proposed projection procedure considers two main steps:

- **Mapping:** The planar (or 2D) images are generated applying the orthographic projection for the different sides (planes) of the precision box, i.e., box surrounding the PC object with a size defined by the coordinates' precision. For each plane, a point is projected onto the plane as long as the point is not occluded by another point closer to the same plane.

- **Filtering:** Points which are projected onto a projection plane but do not belong to the PC surface closer to the plane must be removed. Considering a typical rendering process, these points will be occluded due to the use of primitives around each point or due to surface reconstruction techniques. These points are unduly projected when there is some empty space between points in the surface closer to the plane and, therefore, points from the opposite side of the object are unduly projected onto this plane. Since these points are not visible after PC rendering is performed, a filtering technique is used to remove these points from the projected images.

#### 1) Mapping

The mapping algorithm projects every point visible from the perspective associated to each specific PC precision box plane as follows:

1. Six images, each corresponding to a projection plane are initialized with a uniform background color, e.g., white. In practice, the background color can be any color (in this case 255 is used) since the background pixels will be later padded and, thus, filled with non-uniform, interpolated values. Six planes are defined, notably PL = $\{xy, xz, yz, \acute{x}\acute{y}, \acute{x}\acute{z}, \acute{y}\acute{z}\}$, the first three with (0,0,0) origin and, the last three, including the opposite point, i.e. $(2^p, 2^p, 2^p)$, where $p$ is the PC coordinates precision; these planes are shown in Fig. 4.

Initially, these images are set to 255 (as mentioned above) according to (2):

$$I_{\text{PL}} = \begin{bmatrix} 255 & \cdots & 255 \\ \vdots & \ddots & \vdots \\ 255 & \cdots & 255 \end{bmatrix}_{2^p \times 2^p \times 6} \quad (2)$$

where $I_{\text{PL}}$ is the projected image associated to any of the planes, PL, e.g., $I_{xy}$ or $I_{\acute{y}\acute{z}}$.

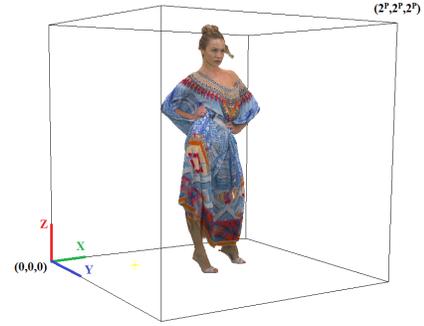

Fig. 4. The six precision box planes surrounding a PC object: $\{xy, xz, yz, \acute{x}\acute{y}, \acute{x}\acute{z}, \acute{y}\acute{z}\}$.

2. Six binary images corresponding to the occupancy map ($OM$) for each projected image are initialized with '0', i.e., non-occupied pixel/voxel, as follows:

$$OM_{PL} = \begin{bmatrix} 0 & \cdots & 0 \\ \vdots & \ddots & \vdots \\ 0 & \cdots & 0 \end{bmatrix}_{2^p \times 2^p \times 6} \quad (3)$$

where $OM_{PL}$ is the occupancy associated to any of the planes PL, e.g., $OM_{xy}$ or $OM_{\acute{y}\acute{z}}$.

3. To keep track of the occluded points, for the three coordinates, two depth maps are used to store the minimum projected depth ($NearD$) and maximum depth ($FarD$). These depth maps are initialized with 0 and $2^p$, respectively, according to (4).

$$NearD = \begin{bmatrix} 0 & \cdots & 0 \\ \vdots & \ddots & \vdots \\ 0 & \cdots & 0 \end{bmatrix}_{2^p \times 2^p}, FarD = \begin{bmatrix} 2^p & \cdots & 2^p \\ \vdots & \ddots & \vdots \\ 2^p & \cdots & 2^p \end{bmatrix}_{2^p \times 2^p} \quad (4)$$

While $NearD$ keeps record of the depth of the closest projected point to the ($\acute{x}\acute{y}, \acute{x}\acute{z}, \acute{y}\acute{z}$) planes, $FarD$ keeps record of the depth of the closest projected point to the ($xy, xz, yz$) planes, for every projected position.

4. For each point $P_i = (p_x, p_y, p_z)$ in the PC with color $C_p$, two parallel planes from the set PL will be jointly processed, starting with ($xy, \acute{x}\acute{y}$). The following steps are performed:

    i.  Define the depth $D_i$ of point $P_i$ as the value for the coordinate $p_z$ since the axis $z$ is the perpendicular axis to both the $xy$ and $\acute{x}\acute{y}$ planes.

    ii. If $D_i$ is less than or equal to the maximum depth at position $(p_x, p_y)$ in $FarD$ for this plane, then the point is projected onto position $(p_x, p_y)$ of image $I_{\acute{x}\acute{y}}$. The corresponding pixel in the occupancy map is also set to '1' and the corresponding maximum depth is updated to the depth $D_i$ as follows:

    if $D_i \leq FarD(p_{\acute{x}}, p_{\acute{y}})$ then $\quad (5)$

    $I_{\acute{x}\acute{y}}(p_{\acute{x}}, p_{\acute{y}}) = C_p$ and

    $OM_{\acute{x}\acute{y}}(p_{\acute{x}}, p_{\acute{y}}) = 1$ and

$$FarD(p_{\acute{x}}, p_{\acute{y}}) = D_i$$

iii.　If $D_i$ is larger than or equal to the minimum depth at position $(p_x, p_y)$ in *NearD* for this plane, then the point is projected onto position $(p_x, p_y)$ of image $I_{x,y}$. The corresponding pixel in the occupancy map is also set to '1' and the corresponding minimum depth is updated to the depth $D_i$ as follows:

if $D_i \geq NearD(p_x, p_y)$ then　　　　　　　(6)

$I_{x,y}(p_x, p_y) = C_p$ and

$OM_{x,y}(p_x, p_y) = 1$ and

$NearD(p_x, p_y) = D_i$

Finally, steps i-iii have to be repeated for the other two pairs of planes in PL, i.e., $(xz, \acute{x}\acute{z})$ and $(yz, \acute{y}\acute{z})$, thus obtaining more four images ($I_{xz}$, $I_{\acute{x}\acute{z}}$, $I_{yz}$, $I_{\acute{y}\acute{z}}$).

### 2) Filtering

At this stage, every point that is not occluded should have been projected. In this context, it is possible that some points from the surface farther to the projection plane may be visible from it and, thus, projected onto it. This occurs because the surfaces close and farther away to the projected plane may not have the same density and may not be aligned. Some pixel positions may be filled from surfaces that are not even visible after rendering (from the perspective of the projected plane).

These points should be filtered out by comparing their depth to the depth of their neighboring pixels in a window $w$ since the depth of these far away points are significantly different from its projected neighboring closest points. The algorithm for filtering the points from the 'back part' of the PC that are not seen by the users from that perspective, proceeds as follows:

1.　For each occupied pixel $(u, v)$ in the projected image for planes $PL \in \{xy, xz, yz\}$, compute the difference between *NearD(u,v)* and the average of *NearD* values for its occupied neighbors in a 2D window with size $w \times w$ centered at $(u, v)$. If this difference is smaller than a predefined positive threshold $\tau$, then set that position to unoccupied in the associated occupancy map and set the corresponding projected image position to the background value as follows:

if $NearD_{\hat{K}}(u, v) - \frac{\sum_{(i,j) \in w} NearD_{PL}(i,j) \times OM_{PL}(i,j)}{\sum_{(i,j) \in w} OM_{PL}(i,j)} \leq \tau$:　　(7)

$I_{PL}(u, v) = 255$ and $OM_{\hat{K}}(u, v) = 0$

2.　For each occupied pixel $(u, v)$ in the projected image for planes $PL \in \{\acute{x}\acute{y}, \acute{x}\acute{z}, \acute{y}\acute{z}\}$, compute the difference between *FarD (u,v)* and the average of *FarD* values for its neighbors in a 2D window with size $w \times w$ centered at $(u, v)$. If this difference is smaller than a predefined positive threshold $\tau$, than set that position to unoccupied in the associated occupancy map and set the corresponding projected image position of the background value as follows:

if $FarD_{\hat{K}}(u, v) - \frac{\sum_{(i,j) \in w} FarD_{PL}(i,j) \times OM_{PL}(i,j)}{\sum_{(i,j) \in w} OM_{PL}(i,j)} \geq \tau$:　　(8)

$I_{PL}(u, v) = 255$ and $OM_{\hat{K}}(u, v) = 0$

The algorithm above is a proximity filtering algorithm that depends on the threshold value $\tau$ that should be set according to the curvature of the objects surface, which is typically not very high. Even if a few PC front points are unduly 'filtered', the impact is small as they will be filled during the padding process performed next. After some experiments, it was found that a fixed filtering threshold $\tau$ of 20 was effective in the filtering of these already projected pixels.

### D. Cropping

The output of the previous step corresponds to six projected images with size $2^p \times 2^p$, one per projection plane. These images may contain a significant amount of background pixels corresponding to empty areas around the PC object projection. To remove the undesired influence of this background data, the excessive background around the projected PC is cropped out to the minimum size including the occupied pixels after projection. First, a bounding box surrounding the PC object(s) for each projected image is obtained for the reference and degraded PCs, using the occupancy map obtained during the projection step. More precisely, by scanning from top left to bottom right, the positions of the first and last occupied pixel in each occupancy map are used to define the bounding box for the object in the projected map. Note that the reference and degraded bounding boxes for each view are identical as using aligned geometries. After, cropping is performed using the obtained bounding box for the reference/degraded projected images associated to each plane.

Fig. 5 shows an example of the cropping operation for the *Longdress* PC. While the full projected image is shown in Fig. 5(a) with the precision bounding box in green, Fig. 5(b) shows the cropped image with a largely reduced background.

### E. Padding

After cropping, the background information is significantly removed. However, background pixels associated to unoccupied points inside the object and some background area around the object still exist. To avoid the negative impact of the background color pixels when computing the 2D quality metric, the color values for the unoccupied pixels (and thus with uniform background values) should be set using some 2D interpolation technique, thus more appropriately filling all empty spaces. This operation aims to replicate the effect of PC rendering, which creates continuous surfaces without holes using appropriate rendering primitives, thus removing any bias on the quality score due to pixels with the same identical value for reference and degraded projected images, i.e., for areas without any distortion. This padding approach allows to apply any 2D quality metric, which uses as input a rectangular image without requiring any adaptations for the 2D quality metric to work with arbitrarily shaped 2D regions, i.e., to consider only foreground pixels. To fill the non-occupied pixels, it is proposed to use an image inpainting technique from the literature called Navier-Strokes [30], which has been selected due to its good performance. The occupancy maps created during the projection process are used as a padding mask to guarantee that only the unoccupied pixels are padded. An example of a padded image is shown in Fig. 5(c).

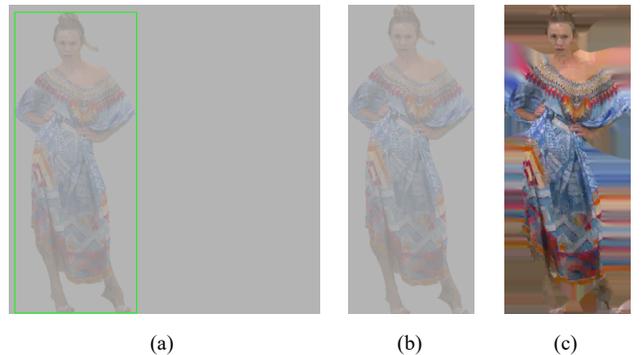

(a)　　　　　　(b)　　　　　　(c)

Fig. 5. Frontal view projection for *Longdress* PC decoded with G-PCC in Octree coding mode and RAHT color coding mode at the medium rate: (a) projected image with the bounding box in green; (b) cropped image; (c) padded image. The padded images look 'brighter' since the non-padded images still have many light grey background pixels inside the foreground area.

### F. Fusion

Before obtaining the final *JGC-ProjQM* PC quality metric, two intermediate scores are computed, *JGC-ProjQM_reference* and *JGC-ProjQM_degraded*, corresponding to the two parallel branches in the architecture, one corresponding to the reference geometry and another to the degraded geometry. To combine the two intermediate quality scores, the following linear model is proposed:

$$JGC\text{-}ProjQM = \alpha\, JGC\text{-}ProjQM_{reference} + \beta\, JGC\text{-}ProjQM_{degraded} \qquad (9)$$

In (14), the $\alpha$ and $\beta$ parameters are estimated using a least squares linear regression procedure that minimizes the residual sum of the squared difference between the objective scores predicted by the linear approximation, *JGC-ProjQM*, and the Mean Opinion Scores (MOS) available in some selected dataset. For this paper, the used dataset was the MPEG Point Cloud Compression Dataset (M-PCCD) [31]. Although more complex models may be selected, they typically require more parameters, bringing the risk of overfitting to the selected dataset. This is rather critical as there are not that many PC datasets available with subjective scores and representative geometry and color degradations, especially compared to image and video datasets.

## IV. PERFORMANCE ASSESSMENT

The main objective of this section is to evaluate the proposed *JGC-ProjQM* PC quality metric compared with the best performing PC quality metrics available in the literature, projection-based or not. Moreover, an ablation study is presented to assess the performance impact of each module in the *JGC-ProjQM* architecture.

### A. Subjective Evaluation Dataset

To perform the performance assessment, the MPEG Point Cloud Compression Dataset (M-PCCD), publicly available in [31], has been selected. This recent dataset includes 232 stimuli where geometry and color are both encoded, which is very suitable to evaluate the proposed *JGC-ProjQM* performance. The M-PCCD dataset includes both the MOS values as well as the reference and degraded/decoded PCs.

The test material in this dataset corresponds to nine PCs, including four objects and five human figures. While *Longdress*, *Loot*, *Redandblack*, *Soldier*, *The20smaria* and *Head* are from the MPEG repository [32], *Romanoillamp* and *Biplane* are from the JPEG repository [33] and *Amphoriskos* from *Sketchfab* dataset [34], see Fig. 6. *Redandback* has been used for training the subjects. The PCs are shown in Fig. 6 and their characteristics listed in Table I.

The PCs have been coded in the following conditions: i) 24 rates for the MPEG G-PCC standard with six different rates for each combination of Octree and TriSoup geometry coding modes with the RAHT and Lifting color coding modes; and ii) five rates for the MPEG V-PCC standard. The rates were selected based on the MPEG Common Test Conditions (CTC) recommendations [15].

The performed subjective quality assessment study has followed the Double Stimulus Impairment Scale (DSIS) methodology and scores have been obtained in two separate labs, each with 20 subjects. A point-based rendering procedure was used, and the PCs were shown side-by-side with an interactive evaluation protocol which allowed subjects to select and modify their viewpoint.

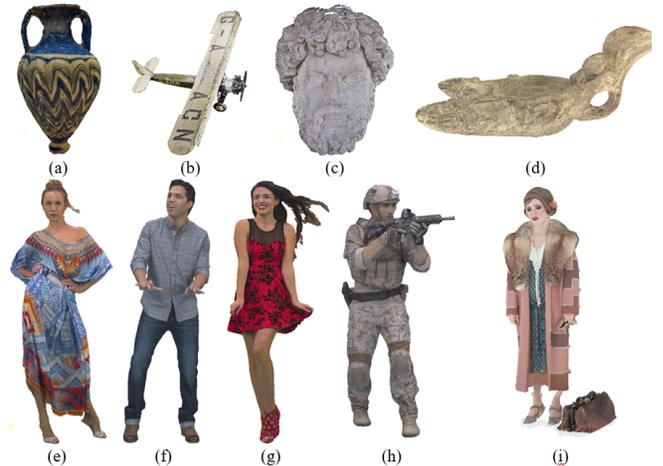

Fig. 6. Test materials in the M-PCCD dataset. From top left to bottom right: (a) *Amphoriskos*; (b) *Biplane*; (c) *Head*; (d) *Romanoillamp*; (e) *Longdress*; (f) *Loot*; (g) *Redandblack*; (h) *Soldier*; (i) *The20smaria*.

TABLE I
M-PCCD TEST PCS AND ASSOCIATED CHARACTERISTICS.

| Name | Type | Repository | Precision | No. Points |
|---|---|---|---|---|
| *Amphoriskos* | Objects | *Sketchfab* | 10-bit | 814.474 |
| *Romanoillamp* | Objects | JPEG repository | 10-bit | 636.127 |
| *Biplane* | Objects | JPEG repository | 10-bit | 1.181.016 |
| *Head* | Objects | MPEG repository | 9-bit | 938.112 |
| *Longdress* | People | MPEG repository | 10-bit | 857.966 |
| *Loot* | People | MPEG repository | 10-bit | 805.285 |
| *Redandblack* | People | MPEG repository | 10-bit | 757.691 |
| *Soldier* | People | MPEG repository | 10-bit | 1.089.091 |
| *The20smaria* | People | MPEG repository | 10-bit | 1.553.937 |

The outlier detection algorithm described in the ITU-R Recommendation BT.500-13 [35] was performed separately for each laboratory, to exclude subjects whose ratings deviated drastically from the rest of the quality scores. As a result, no outliers were identified, thus, leading to 20 ratings per stimulus at each lab. Then the MOS was computed by averaging all the subject scores for each stimulus.

### B. Fusion Parameters Overfitting Checking

The final *JGC-ProjQM* metric is a linear combination of the intermediate quality metrics associated to the two architectural branches as shown in Fig. 1. The $\alpha$ and $\beta$ parameters are estimated using a least squares linear regression procedure which uses the objective and subjective scores for all point clouds and therefore is important to confirm that no overfitting happens. For this purpose, the M-PCCD dataset is randomly split into 75% training data and 25% test data for 100 times. For each iteration, the $\alpha$ and $\beta$ parameter values are estimated from the training data split and used to compute the final PC quality metric score for the test data split. The subjective-objective performance is measured with the Pearson Linear Correlation Coefficient (PLCC) which has been computed for each iteration and is shown in Fig. 7. The average performance for all iterations (i.e., for different splits) is computed as the average PLCC over all iterations and is also shown in Fig. 7 as a red line. The PLCC performance computed using all data for both training and testing is also shown in Fig. 7. The analysis of the results shows that the final *JGC-ProjQM* values are very close to the average *JGC-ProjQM* value, thus implying that the obtained performance is not due to overfitting. This overfitting analysis is made for four 2D quality metrics, notably SSIM, MS-SSIM, FSIM and Y-PSNR.

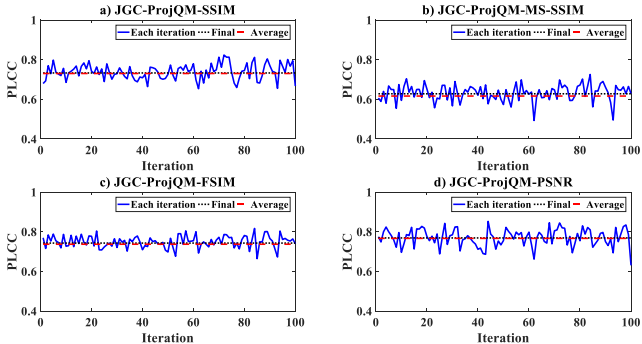

Fig. 7. *JGC-ProjQM* PLCC overfitting analysis for the following 2D quality metrics: (a) SSIM; (b) MS-SSIM; (c) FSIM; (d) Y-PSNR.

### C. Experimental Results and Analysis

In this section, the objective-subjective correlation performance of the proposed PC quality metric[1] is evaluated and compared with some relevant benchmarks. In the following, non-linear fitting is applied to all objective quality scores, in this case, using the following logistic function:

$$MOS_p = \beta_2 + \frac{\beta_1 - \beta_2}{1 + e^{-\left(\frac{Q_i - \beta_3}{\beta_4}\right)}} \qquad (10)$$

[1] A python implementation is made available online at https://github.com/AlirezaJav/Projection-based-PC-Quality-Metric

where $Q_i$ are the objective metric scores and $\beta_1, ..., \beta_4$ are the regression model parameters. This approach allows to fit the objective metric scores to the perceptual (MOS) scale to obtain the fitted predicted MOS scores. To assess the quality metrics performance, the PLCC, Spearman Rank Order Correlation Coefficient (SROCC) and Root Mean Squared Error (RMSE) are used. When the PLCC and SROCC scores are close to '1', the predicted objective quality scores are highly correlated and have a monotonic relationship with the ground-truth MOS scores. As a measure of monotonicity, SROCC does not depend on the selected fitting function. A wide range of objective PC quality metrics available in the literature were selected as benchmarks to evaluate the proposed PC quality metric more effectively. These metrics are listed in Table II.

#### 1) Projection-based Metrics Performance Comparison

In this section, the proposed *JGC-ProjQM* metric performance is compared with the projection-based metrics proposed in [10] and [11]. For a fair comparison, the same 2D quality metric is used in the *JGC-ProjQM* quality metric and the selected benchmarks. In [10], VIFP, SSIM, MS-SSIM and Y-PSNR are used, while in [11], only Y-PSNR is used.

TABLE II
PC QUALITY METRICS SELECTED AS BENCHMARKS FOR PERFORMANCE COMPARISON.

| Metric Name | Short Description |
|---|---|
| D1 and D1-PSNR [15] | Po2Po MSE error and associated PSNR. |
| D2 and D2-PSNR [15] | Po2Pl MSE error and associated PSNR. |
| Hausdorff distance and Hausdorff PSNR [13] | Po2Po and Po2Pl Hausdorff distance and associated PSNR. |
| Y-MSE and Y-PSNR [15] | MSE between luminance of points and their nearest neighbor and associated PSNR. |
| Angle-MSE [16] | Mean squared of angular similarity between underlying surfaces at each point and its nearest neighbor. |
| PCQM [26] | Weighted average of geometry curvature features and lightness, chroma and hue features between reference and decoded PCs. |
| PointSSIM [28] | Color features are extracted in a local neighborhood around each point in the reference and degraded PCs using variance as statistical dispersion measure. |
| $H_{YL}^V$ [21] | Computed from the color histogram and correlogram of the luminance component. |
| $d_{gc}$ [21] | Linear combination of Po2Pl MSE with $H_{YL}^V$. |
| $PCM_{RR}$ [27] | Reduced reference quality metric corresponding to the weighted average of 7 geometry, 7 color and 7 normal features. |
| Geotex [22] | Local Binary Pattern descriptors applied to the luminance are extracted and used for quality metric computation. |
| Proj-Y-PSNR [11] | Projection-based quality metric using six faces of a cube; the six images are concatenated and Y-PSNR are computed. |
| Proj-Y-MSSIM and Proj-Y-VIFP [10] | Projection-based metric using 42 views; the final metric is obtained using several 2D metrics and pooled using l1-norm. |
| GH-PSNR [17] | Generalized Hausdorff distance-based PSNR, considering 98% of the distances, using maximum pooling function. |
| RA-PSNR [19] | Resolution-Adaptive PSNR considering the rendering resolution computed over the 10 nearest neighbors. |
| MMD and MMD-PSNR [18] | PSNR based Point-to-Distribution metric where the mean Mahalanobis distances between a point and a distribution of points is computed. |

TABLE IV
OBJECTIVE-SUBJECTIVE CORRELATION PERFORMANCE COMPARISON OF THE PROPOSED METRIC WITH [10] AND [11] FOR THE SAME 2D QUALITY METRIC.

| Projection-based PC quality metric | 2D quality metric | PLCC | SROCC | RMSE |
|---|---|---|---|---|
| *JGC-ProjQM* | VIFP | **85.5** | **83.0** | **0.760** |
| *Alexiou et al.* [10] | | 74.2 | 71.5 | 0.951 |
| *Gain* | | 11.3 | 11.5 | 0.191 |
| *JGC-ProjQM* | SSIM | **81.3** | **80.9** | **0.800** |
| *Alexiou et al.* [10] | | 63.3 | 62.6 | 1.061 |
| *Gain* | | 18.0 | 18.3 | 0.261 |
| *JGC-ProjQM* | MS-SSIM | **82.8** | **79.5** | **0.830** |
| *Alexiou et al.* [10] | | 75.2 | 70.9 | 0.959 |
| *Gain* | | 7.6 | 8.6 | 0.129 |
| *JGC-ProjQM* | Y-PSNR | **79.1** | **77.1** | **0.870** |
| *Alexiou et al.* [10] | | 62.8 | 66.7 | 1.013 |
| *Gain* | | 16.3 | 10.4 | 0.143 |
| *de Queiroz et al.* [11] | | 33.1 | 43.0 | 1.228 |
| *Gain* | | 46.0 | 34.1 | 0.358 |

TABLE III
OBJECTIVE-SUBJECTIVE CORRELATION PERFORMANCE FOR *JGC-PROJQM* USING DIFFERENT 2D QUALITY METRICS, FROM BEST TO WORST PLCC.

| 2D Metric | PLCC | SROCC | RMSE |
|---|---|---|---|
| *DISTS* | **94.7** | **95.6** | **0.439** |
| *LPIPS* | 92.3 | 93.2 | 0.523 |
| *FSIM* | 88.2 | 90.1 | 0.640 |
| *VSI* | 85.4 | 87.6 | 0.707 |
| *HaarPSI* | 84.8 | 87.7 | 0.721 |
| *VIPF* | 83.0 | 85.5 | 0.758 |
| *SSIM* | 80.9 | 81.3 | 0.800 |
| *MS-SSIM* | 79.5 | 82.8 | 0.825 |
| *PSNR HVS M* | 78.7 | 81.3 | 0.840 |
| *PSNR HVS* | 78.4 | 80.5 | 0.845 |
| *Y-PSNR* | 77.1 | 79.1 | 0.866 |

Table III clearly shows that proposed *JGC-ProjQM* metric significantly outperforms the already available projection-based PC quality metrics, for the same 2D quality metric. The maximum *JGC-ProjQM* gains are 18.0% for PLCC, 18.3% for SROCC and 0.26 for RMSE, comparing to *Alexiou et al.* [10]. These gains are rather large and consistent across the used quality metric performance measures, i.e., PLCC, SROCC and RMSE, especially considering that [10] uses 42 views, which is a much larger (and also more complex) number of views than the six views considered in *JGC-ProjQM*. Other interesting conclusion is that VIFP is the best 2D quality metric and Y-PSNR leads to the worst correlation performance.

*2) 2D Quality Assessment Metrics Performance Impact*

As stated in Section III, the proposed *JGC-ProjQM* metric is flexible enough to accommodate any 2D quality metric. In this section, the performance of the proposed metric is evaluated for several 2D quality metrics. This will allow to identify which 2D quality metric leads to the best correlation performance. In this case, the same 2D quality metric is used in both architectural branches of the proposed quality metric, i.e., reference and degraded, especially because the fusion module works best when the quality range and scale is similar for both the reference and degraded geometry branches. Since the used 2D quality metric can significantly influence the *JGC-ProjQM* performance, a wide set of available quality 2D metrics are evaluated, notably: 1) Y-PSNR, 2) PSNR-HVS [36]; 3) PSNR-HVS-M [37]; 4) Structural Similarity Index Metric (SSIM) [38]; 5) Multi-Scale Structural Similarity Index Metric (MS-SSIM) [39], 6) Visual Information Fidelity Measure (VIFP) [40]; 7) Feature Similarity Index (FSIM) [41]; 8) Visual Saliency Index (VSI) [42]; 9) Learned Perceptual Image Patch Similarity (LPIPS) [43]; 10) Deep Image Structure and Texture Similarity (DISTS) [44]; and 11) Haar Perceptual Similarity Index (HaarPSI) [45].

Table IV shows the *JGC-ProjQM* correlation performance for a large set of 2D quality metrics, considering all possible coding degradations (all codecs data). The DISTS 2D quality metric has the best correlation performance among all the 2D quality metrics while LPIPS, FSIM, and HaarPSI come in the following positions. Both DISTS and LPIPS are very recent 2D quality metrics that use powerful deep-learning features to perform quality assessment. More specifically, DISTS includes both color and structure similarity components, which are weighted to achieve a better correlation with the perceived quality and to be invariant to small changes in color patches (homogenous regions with repeated elements). LPIPS computes distances between features extracted from the reference and degraded projected images at different layers of a neural network. For both DISTS and LPIPS metrics, a perceptual feature space is used. Typically, these quality metrics weight more general appearance changes than small color changes, where their elements may have different location, size, color and orientation. This fits rather well the projected images obtained by the proposed *JGC-ProjQM* metric where small color changes may occur due to the recoloring process. The proposed *JGC-ProjQM* metric with the best four 2D quality metrics, i.e., *DISTS, LPIPS, FSIM, VSI*, will be used for the remaining experiments.

*3) Overall Point Cloud Quality Metrics Performance Comparison*

In this section, the performance of the proposed *JGC-ProjQM* metric is compared with many state-of-the-art PC quality metrics available in the literature. Table V shows the SROCC, PLCC and RMSE scores for the many benchmark quality metrics, using all the codecs scores together as well as the V-PCC and G-PCC scores individually. When a large set of metrics is proposed in a reference, only the best variants are included. This separation can be justified by the fact that the PC quality metrics performance may be significantly influenced by the different type of coding artifacts generated by different PC coding solutions (which may be rather different). This split was performed as follows: i) G-PCC decoded PCs, including TriSoup and Octree geometry coding modes as well as RAHT and Lifting color coding modes; ii) V-PCC decoded PCs; and iii) all decoded PCs together. From Table V, where the best results are highlighted in bold and second best in grey shaded cells, the following conclusions may be derived:

- **Overall correlation performance:** The proposed *JGC-ProjQM* metric using DISTS is the best performing metric for PC quality assessment, achieving the best PLCC, SROCC and RMSE scores. This result also confirms that by projecting a PC into several 2D images (which are close to what a user sees) and after exploiting the power of 2D quality metrics, a top correlation performance can be obtained. The proposed *JGC-ProjQM* with LPIPS and the PointSSIM quality metrics also have a very high correlation performance.

TABLE V
OBJECTIVE-SUBJECTIVE CORRELATION PERFORMANCE COMPARISON BETWEEN THE PROPOSED *JGC-PROJQM* METRIC AND BENCHMARKING METRICS.
CELLS WITH A DASH (-) SHOW MISSING PERFORMANCE SINCE SOME OF THESE RESULTS WERE OBTAINED FROM THE RELEVANT PAPER.

| Type | | Metric Name | All | | | V-PCC | | | G-PCC | | |
|---|---|---|---|---|---|---|---|---|---|---|---|
| | | | PLCC | SROCC | RMSE | PLCC | SROCC | RMSE | PLCC | SROCC | RMSE |
| Point-based | Po2Po | D1 [15] | 84.8 | 86.8 | 0.722 | 46.3 | 42.0 | 0.928 | 88.6 | 90.0 | 0.651 |
| | | D1-PSNR [15] | 77.7 | 79.7 | 0.857 | 30.4 | 28.2 | 0.997 | 82.5 | 83.9 | 0.794 |
| | | Hausdorff [13] | 1.4 | 37.0 | 1.360 | 14.7 | -17.5 | 1.047 | 5.3 | 54.4 | 1.404 |
| | | Hausdorff PSNR [13] | 66.1 | 36.6 | 1.021 | 27.1 | -14.9 | 1.008 | 76.0 | 53.3 | 0.912 |
| | | GH-PSNR [17] | 84.6 | 86.9 | 0.726 | 57.8 | 57.3 | 0.854 | 88.5 | 89.9 | 0.653 |
| | | RA-PSNR [19] | 88.8 | 90.2 | 0.626 | 68.9 | 67.3 | 0.759 | 91.0 | 91.8 | 0.584 |
| | | Y-MSE [15] | 66.3 | 66.2 | 1.018 | 37.9 | 33.3 | 0.969 | 70.3 | 70.3 | 0.998 |
| | | Y-PSNR [15] | 67.1 | 66.2 | 1.009 | 37.6 | 33.3 | 0.970 | 71.4 | 70.3 | 0.984 |
| | Po2Pl | D2 [15] | 85.9 | 88.4 | 0.695 | 73.5 | 68.8 | 0.710 | 87.9 | 90.6 | 0.669 |
| | | D2-PSNR [15] | 80.5 | 83.8 | 0.808 | 60.3 | 55.3 | 0.835 | 83.4 | 87.3 | 0.774 |
| | | Hausdorff [13] | 67.2 | 50.5 | 1.007 | 23.8 | 12.8 | 1.017 | 78.4 | 66.3 | 0.871 |
| | | Hausdorff PSNR [13] | 56.3 | 49.3 | 1.124 | 28.6 | 13.5 | 1.003 | 68.7 | 65.3 | 1.020 |
| | | GH-PSNR [17] | 84.3 | 87.9 | 0.731 | 75.1 | 71.2 | 0.691 | 87.5 | 91.0 | 0.680 |
| | | RA-PSNR [19] | 88.9 | 89.9 | 0.622 | 79.9 | 76.9 | 0.629 | 90.3 | 91.5 | 0.604 |
| | Po2D | MMD [18] | 86.9 | 88.7 | 0.672 | 71.8 | 69.0 | 0.729 | 88.8 | 90.3 | 0.647 |
| | | MMD-PSNR [18] | 86.9 | 88.7 | 0.673 | 71.9 | 69.0 | 0.728 | 88.7 | 90.3 | 0.648 |
| | Pl2Pl | Angle-MSE [15] | 62.4 | 47.7 | 1.063 | 51.6 | 34.1 | 0.897 | 69.0 | 55.0 | 1.016 |
| Feature-based | | PCQM [26] | 89.9 | 91.6 | 0.597 | - | - | - | - | - | - |
| | | $d_{qc}$ [21] | 90.4 | 92.0 | 0.585 | 75.3 | 74.0 | 0.689 | 92.5 | 93.9 | 0.533 |
| | | $H_{L2}^{L1}$ [21] | 85.3 | 88.4 | 0.710 | 65.7 | 68.3 | 0.789 | 87.9 | 91.9 | 0.669 |
| | | $PCM_{RR}$ [27] | 90.2 | 90.7 | 0.573 | 71.6 | 64.8 | 0.731 | 89.2 | 91.0 | 0.636 |
| | | PointSSIM [28] | 92.6 | 91.8 | 0.514 | 83.0 | 84.5 | 0.584 | 94.4 | 92.9 | 0.462 |
| | | Geotex [22] | - | 87.9 | - | - | - | - | - | - | - |
| Projection-based | | Proj-Y-MS-SSIM [10] | 70.9 | 75.2 | 0.959 | 31.9 | 35.4 | 0.992 | 75.3 | 50.1 | 0.924 |
| | | Proj-Y-VIFP [10] | 71.5 | 74.2 | 0.951 | 43.7 | 35.6 | 0.942 | 75.0 | 79.2 | 0.929 |
| | | Proj-Y-PSNR [11] | 43.0 | 33.1 | 1.228 | 29.7 | -10.1 | 1.000 | 47.7 | 35.7 | 1.234 |
| Proposed Projection-based | | JGC-ProjQM-FSIM | 88.2 | 90.1 | 0.640 | 71.0 | 72.1 | 0.737 | 90.3 | 92.1 | 0.604 |
| | | JGC-ProjQM-VSI | 85.4 | 87.6 | 0.707 | 64.5 | 63.9 | 0.809 | 88.1 | 90.1 | 0.664 |
| | | JGC-ProjQM-LPIPS | 92.3 | 93.2 | 0.523 | 80.7 | 79.5 | 0.618 | 93.5 | 94.2 | 0.497 |
| | | JGC-ProjQM-DISTS | 94.7 | 95.6 | 0.439 | 86.4 | 85.3 | 0.526 | 95.8 | 96.0 | 0.402 |

- **JGC-ProjQM vs point-based PC quality metrics:** The *JGC-ProjQM* metric significantly outperforms the point-based D1-PSNR and D2-PSNR and plane-to-plane quality metrics that are currently used by the MPEG and JPEG standardization groups. The *JGC-ProjQM-DISTS* gains over these metrics are rather significant, notably up to 32.3% for PLCC and 47.7% for SROCC for all data. Moreover, *JGC-ProjQM-DISTS* outperforms the best point-based metric in the literature (Po2Po RA-PSNR) by 5.9% for PLCC and 5.4% for SROCC for all data, and by larger gains for all remaining point-based quality metrics.

- **JGC-ProjQM vs feature-based PC quality metrics:** The overall correlation performance of the best proposed projection-based metric, i.e., *JGC-ProjQM-DISTS*, is almost 2.1% in PLCC and 3.8% in SROCC higher than the best feature-based quality metric, i.e., PointSSIM. The feature-based quality metrics often come in second place, achieving also rather good performance, especially compared to point-based quality metrics.

- **V-PCC decoded data:** The *JGC-ProjQM-DISTS* is the best performing metric for V-PCC decoded data with 86.4% for PLCC and 85.3% for SROCC. The benchmark point-based and projection-based quality metrics fail to reliably assess the V-PCC decoded quality. The *JGC-ProjQM-DISTS* metric outperforms the best projection-based metric in the literature (Proj-Y-VIFP) by 42.7% for PLCC and 49.7% for SROCC. For the best point-based metric (Po2Pl RA-PSNR), the performance increase is 6.5% and 8.4% for PLCC and SROCC, respectively. Nowadays, the feature-based metrics are the best performing PC quality metrics in the literature and thus the gains are smaller, 3.4% for PLCC and 0.8% for SROCC, when compared to the feature-based PointSSIM metric. Both, *JGC-ProjQM-DISTS* and PointSSIM show higher performance compared to the other quality metrics.

- **G-PCC decoded data:** The *JGC-ProjQM-DISTS* metric is the best performing metric for G-PCC decoded data with 95.8% for PLCC and 96% for SROCC. Feature-based and point-based metrics (except for Pl2Pl) also show acceptable performance for G-PCC decoded data. However, *JGC-ProjQM-DISTS* outperforms the best MPEG/JPEG adopted metrics (D2-PSNR) by 12.4% for PLCC and 8.7% for SROCC. The correlation gains against the best feature-based quality metric (PointSSIM) are 1.4% for PLCC and 3.1% for SROCC. The benchmark projection-based metrics do not show an acceptable performance for the quality assessment of G-PCC decoded PCs.

TABLE VI
OBJECTIVE-SUBJECTIVE CORRELATION PERFORMANCE FOR THE ABLATION
STUDY OF THE PROPOSED *JGC-PROJQM* METRIC

| Proposed JGC-ProjQM | Correlat. Metric | Full Performance | Removed Module | | |
|---|---|---|---|---|---|
| | | | Recoloring | Cropping | Padding |
| JGC-ProjQM-DISTS | PLCC | 94.7 | 93.7 | 83.7 | 91.4 |
| | SROCC | 95.6 | 94.9 | 86.7 | 91.1 |
| JGC-ProjQM-LPIPS | PLCC | 92.3 | 89.5 | 80.0 | 81.1 |
| | SROCC | 93.2 | 90.7 | 81.8 | 80.8 |
| JGC-ProjQM-FSIM | PLCC | 88.2 | 78.5 | 81.5 | 86.6 |
| | SROCC | 90.1 | 81.5 | 83.7 | 87.5 |
| JGC-ProjQM-VSI | PLCC | 85.4 | 74.6 | 81.6 | 82.7 |
| | SROCC | 87.6 | 77.3 | 83.9 | 83.4 |

### 4) JGC-ProjQM Metric Ablation Study

The proposed *JGC-ProjQM* metric includes several modules that have different impact on the overall correlation performance. To individually assess the impact of each *JGC-ProjQM* module, an ablation study is performed using the entire dataset. More precisely, the *JGC-ProjQM* metric performance is measured for all stimuli (all codecs scores) included in the dataset, each time turning off one of the architectural modules while keeping the others, notably recoloring, cropping, and padding. Table VI shows the PLCC and SROCC results after non-linear logistic fitting for this ablation study.

The following conclusions can be derived:

- **Recoloring**: The correlation performance results show the importance of the recoloring module for the *JGC-ProjQM* performance. For example, for VSI and FSIM, the absence of recoloring leads to losses of 10.8% and 9.7% for PLCC and 10.3% to 8.6% for SROCC. The performance losses after removing recoloring are lower for DISTS and LPIPS, mainly because these two recent 2D quality metrics are robust to geometry distortions and transformations. However, the performance gains by using the recoloring module are up to 10.8% for PLCC and 10.3% for SROCC.

- **Cropping**: The cropping module significantly improves the *JGC-ProjQM* performance by removing background pixels, around the projected PC, that are common to the reference and degraded PCs. These background areas work as a distractor for the 2D quality assessment metric and lower the *JGC-ProjQM* prediction power, even when the background pixels are padded. The performance gains associated to the cropping module are up to 12.3% for PLCC and 11.4% for SROCC.

- **Padding**: The padding module also improves the *JGC-ProjQM* correlation performance. While most of the excessive background is removed by cropping, some background area around the object remains. Moreover, background pixels that are visible in the object surface, often due to sparse sampling during acquisition or the removal of points during coding, are filled by padding. The performance gains by using the padding module are up to 11.2% for PLCC and 12.4% for SROCC.

It is important to note that the correlation performance gains for these modules may be significantly higher when less powerful 2D image metrics, i.e., with lower overall correlation performance, are used, such as MS-SSIM.

## V. CONCLUSIONS

This paper proposes a novel joint geometry and color projection-based PC quality metric, *JGC-ProjQM,* that compares PCs in the 2D domain for two geometry conditions, i.e., reference and degraded geometry. The projection-based PC quality metric applies a projection to obtain six projected images, corresponding to different views over the PC, which are after cropped and padded before performing a 2D quality assessment. After, any 2D quality metric can be applied and the intermediate quality scores for the two geometry conditions are fused to obtain the final *JGC-ProjQM* quality score. The objective-subjective correlation results show that proposed *JGC-ProjQM* metric outperforms all the state-of-the-art PC quality metrics and, thus, PC quality can be efficiently measured in the 2D domain, especially when powerful 2D quality metrics are also exploited. As future work, some visual saliency information and attention models could be integrated in the proposed PC quality assessment framework, thus further improving the correlation performance.